\documentclass[envcountsect,envcountsame]{llncs}
\usepackage{makeidx}
\usepackage{amsfonts}
\usepackage{amsmath}
\usepackage{amssymb}
\usepackage{amscd}
\usepackage[dvips]{graphicx}
\usepackage{algorithmic}
\usepackage{algorithm}
\usepackage{comment}
\usepackage{color}
\usepackage{array,amssymb,amsmath,amsbsy}
\usepackage{enumerate}
\usepackage[bottom]{footmisc}
\usepackage[all,cmtip]{xy}
\usepackage{hyperref}
\usepackage{compress}

\makeatletter
\renewcommand*\l@author[2]{}
\renewcommand*\l@title[2]{}
\makeatletter
\newcommand{\nocontentsline}[3]{}
\newcommand{\tocless}[2]{\bgroup\let\addcontentsline=\nocontentsline#1{#2}\egroup}

\newenvironment{packed_enum}{
    \begin{enumerate}
        \setlength{\itemsep}{1pt}
        \setlength{\parskip}{0pt}
        \setlength{\parsep}{0pt}
}{\end{enumerate}}

\newenvironment{packed_itemize}{
    \begin{itemize}
        \setlength{\itemsep}{1pt}
        \setlength{\parskip}{0pt}
        \setlength{\parsep}{0pt}
}{\end{itemize}}

\newcommand{\heading}[1]{\smallskip\par\noindent{\bf #1}}
\def\claimqed{\hfill$\diamond$}

  \def\calC{{\cal C}}

   \def\calP{{\cal P}}

   \def\frakD{{\mathfrak D}}

\def\frakM{{\mathfrak M}}

\def\frako{{\mathfrak o}}

\def\cP{\hbox{\rm \sffamily P}}
\def\cNP{\hbox{\rm \sffamily NP}}
\def\cGI{\hbox{\rm \sffamily GI}}

\def\O{\mathcal{O}{}}

\def\MD{\frakM\frakD}

\def\int{\hbox{\rm \sffamily INT}}

\def\comp{\hbox{\rm \sffamily COMP}}
\def\cocomp{\hbox{\rm \sffamily co-COMP}}

\def\fun{\hbox{\rm \sffamily FUN}}
\def\perm{\hbox{\rm \sffamily PERM}}

\def\tree{\hbox{\rm \sffamily TREE}}

\def\dim#1{\hbox{\sffamily $#1$-DIM}}

\def\Aut{{\rm Aut}}

\def\sym{\mathbb{S}}

\def\cyc{\mathbb{Z}}

\def\dm{{\rm dim}}

\spnewtheorem*{observation*}{Observation}{\bfseries}{\rmfamily}

\newcounter{lth}
\title{Automorphism Groups of Comparability Graphs\thanks{%
		Supported by CE-ITI (P202/12/G061 of GA\v{C}R) and 
		Charles University as
		GAUK 196213.}}
\author{Pavel Klav\'{\i}k
        \and Peter Zeman}

\institute{Computer Science Institute, Charles University in Prague,\\
            Czech Republic. E-mails: \texttt{\{klavik,zeman\}@iuuk.mff.cuni.cz}.}

\begin{document}
\maketitle

\begin{abstract}
Comparability graphs are graphs which have transitive orientations.  The
dimension of a poset is the least number of linear orders whose intersection
gives this poset.  The dimension $\dm(X)$ of a comparability graph $X$ is the
dimension of any transitive orientation of $X$, and by \dim{k} we denote the
class of comparability graphs $X$ with $\dm(X) \le k$.  It is known that the
complements of comparability graphs are exactly function graphs and permutation
graphs equal \dim2.

\hskip 1em In this paper, we characterize the automorphism groups of permutation
graphs similarly to Jordan's characterization for trees (1869). For permutation
graphs, there is an extra operation, so there are some extra groups not realized
by trees. For $k \ge 4$, we show that every finite group can be realized as the
automorphism group of some graph in \dim{k}, and testing graph isomorphism for
\dim{k} is \cGI-complete.
\end{abstract}

\section{Introduction}

\begin{figure}[t]
\centering
\includegraphics[scale=0.8]{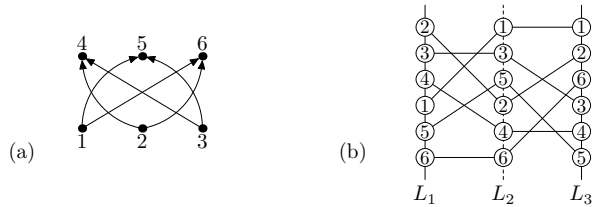}
\caption{(a) A comparability graph with one of its transitive orientations. (b) A
function representation of its complement constructed using three linear orders.}
\label{fig:comparability_example}
\end{figure}

\heading{Comparability Graphs.}
A \emph{comparability graph} is created from a poset by removing the orientation of the edges.
Alternatively, every comparability graph $X$ can be transitively oriented: if $x \to y$ and $y \to
z$, then $xz \in E(X)$ and $x \to z$; see Fig.~\ref{fig:comparability_example}a. This class was
first studied by Gallai~\cite{gallai1967transitiv} and we denote it by \comp.

An important parameter of a poset $P$ is its \emph{Dushnik-Miller
dimension}~\cite{dushnik_miller}.  It is the least number of linear orderings $L_1,\dots,L_k$ such
that $P = L_1 \cap \cdots \cap L_k$.  (For a finite poset $P$, its dimension is always finite since
$P$ is the intersection of all its linear extensions.) Similarly, we define the \emph{dimension} of
a comparability graph $X$, denoted by $\dm(X)$, as the dimension of any transitive orientation of $X$.
(It can be shown that every transitive orientation has the same dimension.) By $\dim{k}$, we denote
the subclass consisting of all comparability graphs $X$ with $\dm(X) \le k$. We get the following
infinite hierarchy of graph classes:
$$\dim1 \subsetneq \dim2 \subsetneq \dim3 \subsetneq \dim 4 \subsetneq \cdots \subsetneq \comp.$$

\heading{Function and Permutation Graphs.} An intersection representation of a graph $X$ is a
collection of sets $\{R_u : u \in V(X)\}$ such that $R_u \cap R_v \ne \emptyset$ if and only if $uv
\in E(X)$; i.e., it encodes the vertices by sets and the edges by intersections of these sets.
To get nice graph classes, one typically restricts the sets $R_v$ to particular classes of
geometrical objects.

\begin{figure}[b]
\centering
\includegraphics[scale=0.8]{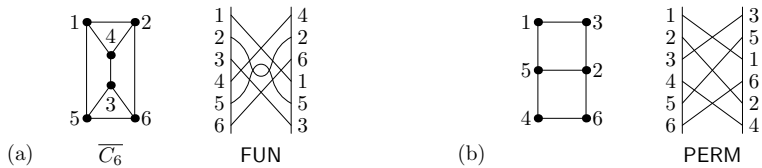}
\caption{(a) A function graph which is not a permutation graph and one of its representations. (b) A
permutation graph and one of its representations.}
\label{fig:fun_perm_example}
\end{figure}

We study the class of \emph{function graphs} (\fun) which are intersection graphs of continuous
functions $R_u : [0,1] \to \mathbb R$ and its subclass \emph{permutation graphs} (\perm) which can
be represented by linear functions~\cite{permutation_graphs}; see Fig.~\ref{fig:fun_perm_example}.

Surprisingly, these classes are related to comparability graphs.
Golumbic~\cite{golumbic1983comparability} proved that function graphs are the complements
of comparability graphs: $\fun = \cocomp$. If two functions do not intersect, we can orient the
non-edge from the bottom function to the top one which gives a transitive orientation of the
complement. On the other hand, a comparability graph has some dimension $k$, so one of
its transitive orientations can be written as $L_1 \cap \cdots \cap L_k$. We place the vertices in
these orderings on $k$ vertical lines between $[0,1]$. Then we represent each vertex by
the polyline function which connects this vertex in each of the $k$ vertical lines; see
Fig.~\ref{fig:comparability_example}b. We get a function representation of the complement. The second
relation $\perm = \comp \cap \cocomp = \dim2$ was shown by Even~\cite{even1972permutation}.

\heading{Automorphism Groups of Graphs.} The automorphism group $\Aut(X)$ of a graph $X$ describes
its symmetries. Every \emph{automorphism} is a permutation of the vertices which preserves
adjacencies and non-adjacencies.  Frucht~\cite{frucht1939herstellung} proved that every finite group
is isomorphic to $\Aut(X)$ of some graph $X$.  Most graphs are asymmetric, i.e., have only the
trivial automorphism~\cite{er63}.  However, many combinatorial and graph theory results rely on
highly symmetrical graphs.

\begin{definition}
For a class $\calC$ of graphs, let $\Aut(\calC)$ be the set $\{\Aut(X) : X \in \calC\}$ of abstract
groups. The class $\calC$ is called \emph{universal} if every abstract finite group is contained in
$\Aut(\calC)$, and \emph{non-universal} otherwise.
\end{definition}

In 1869, Jordan~\cite{jordan1869assemblages} characterized the automorphism groups of trees (\tree).
The automorphism groups of planar graphs were characterized by Babai~\cite{babai1972automorphism};
see also~\cite{FKKN14}. Several results for the automorphism groups of intersection-defined
classes of graphs were shown recently by Klav\'{\i}k and Zeman~\cite{kz}: the automorphism groups of
interval graphs (\int) are the same as of trees, the automorphism groups of unit interval graphs are
the same as of disjoint unions of caterpillars and the automorphism groups of circle graphs are the
same as of pseudoforests; see~\cite{kz} for definitions of these classes.  Most superclasses are
already universal, e.g., chordal graphs, function graphs, claw-free graphs.

\heading{Graph Isomorphism Problem.} This famous problem asks whether two input graphs $X$ and $Y$
are the same up to a relabeling. It obviously belongs to \cNP, and it is not known to be
polynomially-solvable or \cNP-complete. This is a prime candidate for
an intermediate problem with the complexity between \cP\ and \cNP-complete.  It belongs to the low
hierarchy of \cNP~\cite{schoning1988graph}, which implies that it is unlikely \cNP-complete.
(Unless the polynomial-time hierarchy collapses to its second level.) It is closely related to
computing generators of an automorphism group: $X$ and $Y$ are isomorphic if and only if there
exists an automorphism swapping them in $X \mathbin{\dot\cup} Y$, and generators of $\Aut(X)$ can be
computed using $\O(n^4)$ instances of graph isomorphism~\cite{mathon1979note}.  By $\cGI$, we denote
the complexity class of all problems that can be reduced to graph isomorphism in
polynomial time.

For many graph classes, the graph isomorphism problem was shown to be polynomial-time solvable. For
classes like interval graphs and planar graphs~\cite{automorphism_algorithms_trees_int_planar},
circle graphs~\cite{hsu1995m} and permutation graphs~\cite{permutation_isomorphism}, using known
structural results their isomorphism can be reduced to isomorphism of trees.  As evidenced
by~\cite{babai1972automorphism,kz} and this paper, their automorphism groups also have nice
structures.  When a class of graphs has very restrictive automorphism groups, it seems that graph
isomorphism problem should be relatively easy to solve. Actually, the complexity of graph
isomorphism testing of asymmetric graphs is unknown. There are also very complicated polynomial-time
algorithms solving graph isomorphism for universal graph classes: graphs of bounded
degree~\cite{luks1982isomorphism} and with excluded topological subgraphs~\cite{grohe_marx}.

There are graph classes for which testing graph isomorphism is \cGI-complete. For instance, it is
\cGI-complete for bipartite graphs: For a graph $X$, we subdivide its edges which makes it bipartite,
alternatively it is the incidence-graph of $V(X)$ and $E(X)$. Notice that $X \cong Y$ if and only if
their subdivisions are. Similar constructions are known for chordal
graphs~\cite{automorphism_algorithms_trees_int_planar} and other graph classes.  We are not aware of
any \cGI-completeness results for classes with very restricted automorphism groups.  When the graph
isomorphism problem is \cGI-complete, it seems that its automorphism groups have to be rich enough to
encode most graphs (not necessary universal). 

\heading{Our Results.} Since \dim1 consists of all complete graphs, $\Aut(\dim1) = \{\sym_n\}$.
Concerning $\dim2 = \perm$, it was observed in~\cite{kz} that it is non-universal since its
superclass circle graphs is non-universal. In this paper, we answer the question of~\cite{kz} and
describe their automorphism groups precisely:

\begin{theorem} \label{thm:perm_aut_groups}
The class $\Aut(\perm)$ is the class of groups closed under
\begin{packed_enum}
\item[(a)] $\{1\} \in \Aut(\perm)$,
\item[(b)] If $G_1,G_2 \in \Aut(\perm)$, then $G_1 \times G_2 \in \Aut(\perm)$.
\item[(c)] If $G \in \Aut(\perm)$, then $G \wr \sym_n \in \Aut(\perm)$.
\item[(d)] If $G_1,G_2,G_3 \in \Aut(\perm)$, then $(G_1^4 \times G_2^2 \times G_3^2) \rtimes
\cyc_2^2 \in \Aut(\perm)$.
\end{packed_enum}
\end{theorem}

In (d), $\cyc_2^2$ acts on $G_1^4$ as on the vertices of a rectangle, on $G_2^2$ as on
centers of two opposite edges, and on $G_3^2$ as on centers of the other two opposite edges.
Our characterization is similar to Jordan's characterization~\cite{jordan1869assemblages} of the
automorphism groups of trees which consists of (a) to (c). Therefore, $\Aut(\tree) \subsetneq
\Aut(\perm)$. 

Inspired by the technique described in~\cite{kz}, we study the induced action of $\Aut(X)$ on the
set of all transitive orientations.  In the case of permutation graphs, we study the action on pairs
of orientations of the graph and its complement, and show that it is semiregular.  They are
efficiently captured by the modular decomposition which we encode into the modular tree.

We are not aware of any algorthmic result for computing automorphism groups of permutation graphs.
From our description, a polynomial-time algorithm can be constructed.  Further, it can give
$\Aut(X)$ in terms of group products of Theorem~\ref{thm:perm_aut_groups} which gives more insight
into the structure of $\Aut(X)$.

Comparability graphs are universal since they contain bipartite graphs; we can orient all edges from
one part to the other. Since the automorphism group is preserved by complementation and $\fun =
\cocomp$, we have $\Aut(\fun) = \Aut(\comp)$ and function graphs are also universal.  We explain
this in more detail using the induced action on all transitive orientations.

It is well-known that general bipartite graphs have arbitrary large dimensions: the \emph{crown
graph}, which is $K_{n,n}$ without a matching, has dimension $n$.  We give a different
construction which encodes any graph $X$ into a comparability graph $Y$ with $\dm(Y) \le 4$.

\begin{theorem} \label{thm:kdim_aut_groups_and_gi}
For every $k \ge 4$, the class $\dim{k}$ is universal and its graph isomorphism
problem is $\cGI$-complete.
\end{theorem}

Yannakakis~\cite{yannakakis1982complexity} proved that recognizing \dim3\ is $\cNP$-complete by a
reduction from $3$-coloring. For each graph $X$, a comparability graph $Y$ with several vertices
representing each element of $V(X) \cup E(X)$ is constructed. It is shown that $\dm(Y) = 3$ if and
only if $X$ is $3$-colorable. Unfortunately, the automorphisms of $X$ are lost in $Y$ since it
depends on the labels of $V(X)$ and $E(X)$ and $Y$ contains some additional edges according to these
labels. We describe a simple completely different construction which achieves only dimension 4, but
preserves the automorphism group: for a given graph $X$, we create $Y$ by replacing each edge with a
path of length eight. However, it is non-trivial to show that $Y \in \dim4$, and the constructed
four linear orderings are inspired by~\cite{yannakakis1982complexity}.

\heading{Outline.} In Section~\ref{sec:modular_decomposition}, we describe the modular decomposition
and modular trees. In Section~\ref{sec:comparability_graphs}, we discuss the action of
$\Aut(X)$ on the set of all transitive orientations of a comparability graph $X$.
In Section~\ref{sec:permutation_graphs}, we describe automorphism groups of permutation graphs.  In
Section~\ref{sec:kdim}, we encode arbitrary graphs into four-dimensional comparability graphs.  We
conclude this paper with open problems.

\heading{Definitions.}
We use $X$ and $Y$ for graphs, $M$ for modules, $T$ for modular trees and $G$, $H$ for
groups. The vertices and edges of $X$ are $V(X)$ and $E(X)$. The complement of $X$ is
$\overline{X}$. A permutation $\pi$ of $V(X)$ is an \emph{automorphism} if $uv \in E(X) \iff
\pi(u)\pi(v) \in E(X)$. $\sym_n$ and $\cyc_n$ are the \emph{symmetric} and \emph{cyclic groups}.

Given two
groups $N$ and $H$, and a group homomorphism $\varphi \colon H \to \Aut(N)$, we
can construct a new group $N \rtimes_{\varphi} H$ as the Cartesian product $N
\times H$ with the operation defined as $(n_1, h_1) \cdot (n_2,h_2) = (n_1
\cdot \varphi(h_1)(n_2), h_1\cdot h_2)$.  The group $N\rtimes_{\varphi}H$ is
called the \emph{semidirect product of $N$ and $H$ with respect to the
homomorphism $\varphi$}. The \emph{wreath product} $G \wr \sym_n$ is a
shorthand for $G^n \rtimes_{\psi} \sym_n$ where $\psi$ is defined naturally by
$\psi(\pi) = (g_1,\dots,g_n) \mapsto (g_{\pi(1)},\dots,g_{\pi(n)})$.

\section{Modular Decomposition}\label{sec:modular_decomposition}

In this section, we introduce the modular decomposition of a graph $X$ and show that it can be
encoded by a modular tree. We further show that the automorphism group of this modular tree is
isomorphic to $\Aut(X)$.

\begin{figure}[b]
\centering
\includegraphics[scale=0.75]{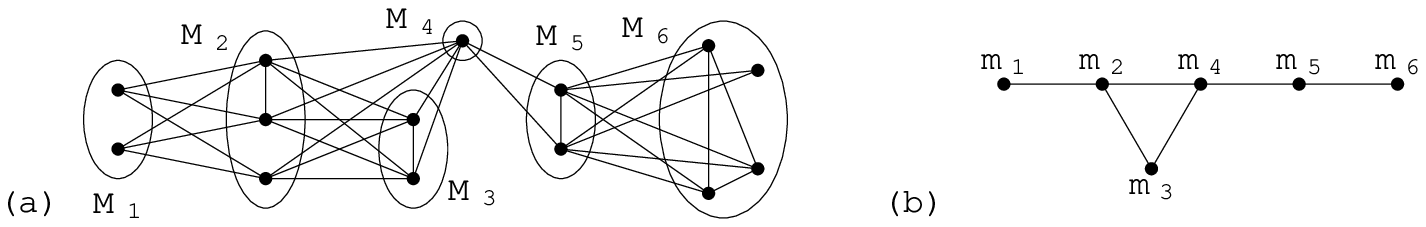}
\caption{(a) A graph $X$ with a modular partition $\calP$ formed by its inclusion maximal
non-trivial modules. (b) The quotient graph $X/\calP$ is prime.}
\label{fig:modules}
\end{figure}

\heading{Modules.}
A \emph{module} $M$ of a graph $X$ is a set of vertices such that each $x \in V(X)\backslash M$ is
either adjacent to all vertices in $M$, or to none of them.  Modules generalize connected
components, but unlike connected components, one module can be a proper subset of another one.
Therefore, modules lead to a recursive decomposition of a graph, instead of just a partition. See
Fig.~\ref{fig:modules}a for examples. A module $M$ is called \emph{trivial} if $M=V(X)$ or $|M|=1$,
and \emph{non-trivial} otherwise.

If $M$ and $M'$ are two disjoint modules, then either the edges between $M$ and $M'$ form the
complete bipartite graph, or there are no edges at all; see Fig.~\ref{fig:modules}a.
In the former case, $M$ and $M'$ are called \emph{adjacent}, otherwise they are
\emph{non-adjacent}.

\heading{Quotient Graphs.}
Let $\calP = \{M_1, \dots, M_k\}$ be a \emph{modular partition} of $V(X)$, i.e., each $M_i$ is a
module of $X$, $M_i \cap M_j = \emptyset$ for every $i\neq j$, and $M_1 \cup \cdots \cup M_k =
V(X)$. We define the \emph{quotient graph} $X/\calP$ with the vertices $m_1,\dots,m_k$ (which
correspond to the modules $M_1,\dots,M_k$) where $m_im_j \in E(X/\calP)$ if and only if $M_i$ and
$M_j$ are adjacent.  In other words, the quotient graph is obtained by contracting each module $M_i$
into a single vertex $m_i$; see Fig.~\ref{fig:modules}b.

\heading{Modular Decomposition.}
We decompose a graph $X$ by finding some modular partition $\calP = \{M_1, \dots, M_k\}$,
computing $X / \calP$ and recursively decomposing $X / \calP$ and each $X[M_i]$. The recursive
process stops on \emph{prime graphs} which are graphs containing only trivial modules.
There might be many such decompositions, depending on the choice of $\calP$ in each
step.  In 1960s, Gallai~\cite{gallai1967transitiv} described the \emph{modular decomposition} in
which some special modular partitions are chosen. This modular decomposition encodes all possible
decompositions.

The key is the following observation. Let $M$ be a module of $X$ and let $M' \subseteq M$. Then $M'$
is a module of $X$ if and only if it is a module of $X[M]$. We construct the modular decomposition
$\MD$ of a graph $X$ in the following way:
\begin{itemize}
\item A graph $X$ is called \emph{degenerate} if it is $K_n$ or $\overline{K}_n$.  If $X$ is a prime
or a degenerate graph, then we add $X$ to $\MD$ and stop.  We stop on degenerate graphs to
make the modular decomposition unique; there are many modular partitions for
them but they are not very
interesting.
\item Let $X$ and $\overline{X}$ be connected graphs. Gallai~\cite{gallai1967transitiv} shows that
the inclusion maximal non-trivial modules of $X$ form a modular partition $\calP$ of $V(X)$, and the
quotient graph $X/\calP$ is a prime graph; see Fig.~\ref{fig:modules}. We add $X/\calP$ to
$\MD$ and recursively decompose $X[M]$ for each $M \in \calP$.
\item If $X$ is disconnected and $\overline{X}$ is connected, then every union of several connected
components is a module. All other modules are subsets of a single connected component. Therefore the
connected components form a modular partition $\calP$ of $V(X)$, and the quotient graph $X/\calP$ is
an independent set. We add $X/\calP$ to $\MD$ and recursively decompose $X[M]$ for each $M
\in \calP$.
\item If $\overline{X}$ is disconnected and $X$ is connected, then the modular decomposition is
defined in the same way on the connected components of $\overline{X}$.  They form a modular
partition $\calP$ and the quotient graph $X/\calP$ is a complete graph. We add $X/\calP$ to
$\MD$ and recursively decompose $X[M]$ for each $M \in \calP$.
\end{itemize}
Gallai~\cite{gallai1967transitiv} shows that the modular decomposition of a graph is unique. It is
easy to see that it captures all modules of $X$.

\heading{Modular Tree.} Let $\MD$ be the modular decomposition of $X$. We encode it by the
\emph{modular tree} $T$ which is a graph with two types of vertices (normal and \emph{marker}
vertices) and two types of edges (normal and directed tree edges). The tree edges connect the
prime and degenerate graphs obtained in $\MD$ into a tree. Further every modular tree has an induced
subgraph called \emph{root node}.   

\begin{figure}[t]
\centering
\includegraphics[scale=0.75]{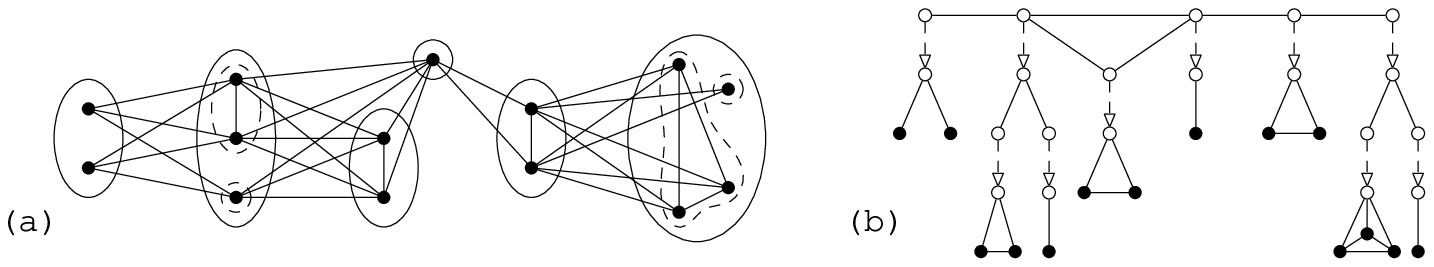}
\caption{(a) The graph $X$ from Fig.~\ref{fig:modules} with the modular partition $\calP$ of $X$ is
depicted, the subsequent modular partitions are depicted by dashed lines. (b) The modular tree $T$
of $X$, the marker vertices are white, the tree edges are dashed.}
\label{fig:modular_tree}
\end{figure}

If $X$ is a prime or a degenerate graph, we define $T = X$ and its root node is equal $T$.
Otherwise, let $\calP = \{M_1,\dots,M_k\}$ be the modular partition of $X$ used in $\MD$ and let
$T_1,\dots,T_k$ be the corresponding modular trees for $X[M_1],\dots,X[M_k]$ according $\MD$. The
modular tree $T$ is constructed by taking disjoint union of $T_1,\dots,T_k$ and the quotient $X /
\calP$ with the marker vertices $m_1,\dots,m_k$. To every graph $T_i$, we add a new marker vertex
$m'_i$ such that $m'_i$ is adjacent exactly to the vertices of the root node of $T_i$. We further
add a tree edge from $m_i$ to $m'_i$.  For an example, see Fig.~\ref{fig:modular_tree}.

Since the modular decomposition of $X$ is unique, also the modular tree of $X$ is unique.
The graphs obtained in $\MD$ are called \emph{nodes of $T$}, or alternatively root nodes of some
modular tree in the construction of $T$.  For a node $N$, its subtree is the modular tree which has
$N$ as the root node.  Every node either has all vertices as marker vertices, or contains no marker
vertices. In the former case, it is called an \emph{inner node}, otherwise a \emph{leaf node}.

An automorphism of the modular tree $T$ has to preserve the types of vertices
and edges.  We denote the automorphism group of $T$ by $\Aut(T)$. For the proof
of the following lemma see Appendix~\ref{sec:app_modular_trees}.

\begin{lemma}\label{lem:aut_group_modular_tree}
If $T$ is the modular tree representing a graph $X$, then
$$\Aut(X) \cong \Aut(T).$$
\end{lemma}

\heading{Recursive Construction.}
We can build $\Aut(T)$ from simple groups recursively, similarly to
Jordan~\cite{jordan1869assemblages}. Suppose that we know automorphism groups
$\Aut(T_1),\dots,\Aut(T_k)$ of all children $T_1,\dots,T_k$ of $T$. Let $R$ be the root node of $T$.
We further color the marker vertices in $R$ by the colors coding isomorphism classes of the subtrees
$T_1,\dots,T_k$, and let $\Aut(R)$ be the color preserving automorphism group of $R$.
Then we get:

\begin{lemma} \label{lem:recursive_aut_modular_tree}
We have
$$\Aut(T) \cong \bigl(\Aut(T_1) \times \cdots \times \Aut(T_k)\bigr) \rtimes \Aut(R).$$
\end{lemma}

\begin{proof}[Sketch]
We isomorphically label the vertices of isomorphic subtrees $T_i$. Each automorphism $\pi \in
\Aut(T)$ is a composition of two automorphisms $\sigma \cdot \tau$ where $\sigma$ maps each subtree
$T_i$ to itself, and $\tau$ permutes the subtrees as in $\pi$ while preserving the labeling.
Therefore, the automorphisms $\sigma$ can be bijectively identified with the elements of the direct
product $\Aut(T_1) \times \cdots \times \Aut(T_k)$ and the automorphisms $\tau$ with some element of
$\Aut(R)$. The rest of the proof follows from a standard argument from
permutation group theory; see Appendix~\ref{sec:app_modular_trees}.\qed
\end{proof}

With no further assumptions on $X$, if $R$ is a prime graph, then $\Aut(R)$ can be isomorphic to an
arbitrary group. If $R$ is a degenerate graph, then $\Aut(R)$ is a direct product of symmetric
groups.

We note that this procedure does not lead to a polynomial-time algorithm for computing $\Aut(T)$.
The reason is that the automorphism groups of prime graphs can be very complicated. To color the
marker vertices, we have to be able to solve graph isomorphism of subtrees $T_i$, and then we have
to find the subgroup of $\Aut(R)$ which preserves the colors.

\section{Automorphism Groups of Comparability Graphs} \label{sec:comparability_graphs}

In this section, we give a structural understanding of the automorphism groups of comparability
graphs, in terms of actions on sets of transitive orientations.

\heading{Structure of Transitive Orientations.} Let $\to$ be a transitive orientation of $X$ and let
$T$ be the modular tree representing $X$.  For modules $M_1$ and $M_2$, we write $M_1 \to M_2$ if
$x_1 \to x_2$ for all $x_1 \in M_1$ and $x_2 \in M_2$.  Gallai~\cite{gallai1967transitiv} shows:
\begin{packed_itemize}
\item If two modules $M_1$ and $M_2$ are adjacent, then either $M_1 \to M_2$, or $M_1 \leftarrow M_2$.
\item The graph $X$ is a comparability graph if and only if each node of $T$ is a comparability
graph.
\item Every prime comparability graph has exactly two transitive orientations, one being the
reversal of the other.
\end{packed_itemize}

The modular tree $T$ encodes all transitive orientations as follows. For each prime node of $T$, we
choose one of the two possible orientations. For each degenerate node, we choose some orientation.
(If it is a complete graph $K_n$, it has $n!$ possible orientations, if it is an independent graph
$\overline{K_n}$, it has the unique orientation).  A transitive orientation of $X$ is then
constructed as follows.  We orient the vertices of leaf nodes as above.  For every subtree with
children $M_1,\dots,M_k$, we orient $X[M_i] \to X[M_j]$ if and only if $m_i \to m_j$ in the root
node.  It is easy to check that this gives a valid transitive orientation, and every transitive
orientation can be constructed in this way.

\begin{figure}[t]
\centering
\includegraphics[scale=0.8]{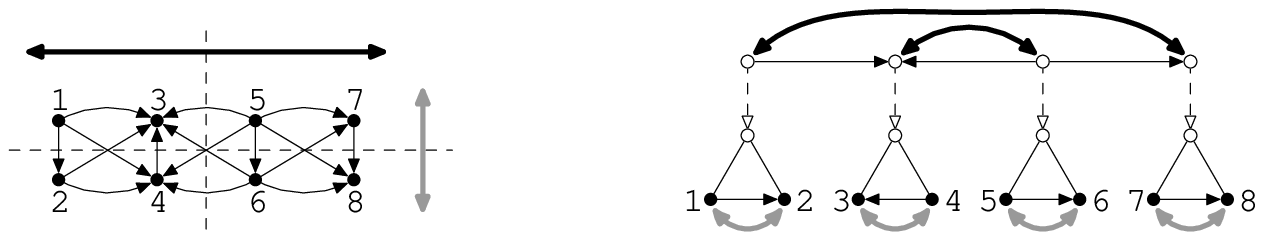}
\caption{Two automorphism reflect $X$ and change the transitive orientation. On the right, their
action on the modular tree $T$.}
\label{fig:comparability_modular_tree}
\end{figure}

\heading{The Induced Action.} Let $\frako(X)$ be the set of all transitive orientations of $X$.  Let
$\pi \in \Aut(X)$ and $O \in \frako(X)$. We define the orientation $\pi(O)$:
$$xOy \iff \pi(x) \pi(O) \pi(y),\qquad \forall x,y \in V(X).$$
We can observe that $\pi(O)$ is a transitive orientation of $X$, so $\pi(O) \in \frako(X)$; see
Fig.~\ref{fig:comparability_modular_tree}. Therefore
$\Aut(X)$ defines an action on $\frako(X)$.

Let $S$ be the stabilizer of some orientation $O$. It consists of all automorphisms which preserve
this orientation, so they permute only the vertices that are incomparable in $O$. In other words,
$S$ is the automorphism group of the poset created by the transitive orientation $O$ of $X$.  We
want to understand it in terms of $\Aut(T)$ for the modular tree $T$ representing $X$. Each
automorphism $\Aut(T)$ somehow acts inside each node, and somehow permutes the nodes, as
characterized in Lemma~\ref{lem:recursive_aut_modular_tree}.

Consider some subtree of $T$ with the children $T_1,\dots,T_k$. Suppose that $\sigma \in S$ maps
$T_i$ to $\sigma(T_i) = T_j$. Then the marker vertices $m_i$ and $m_j$ have to be incomparable in the
root node of this subtree. If the root node is an independent set, the isomorphic subtree can be
arbitrarily permuted in $S$. If it is a complete graph, all subtrees are preserved in $S$. If it is
a prime graph, then isomorphic subtrees of incomparable marker vertices can be permuted.

%

%
%
%
%

\section{Automorphism Groups of Permutation Graphs} \label{sec:permutation_graphs}

In this section, we prove the characterization of the automorphism groups of permutation graphs of
Theorem~\ref{thm:perm_aut_groups}.

\begin{figure}[t]
\centering
\includegraphics[scale=0.8]{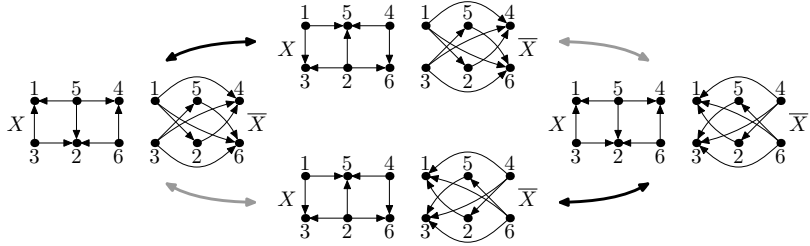}
\caption{The action of $\Aut(X)$ on four pairs of transitive orientations $X$. The black generator
flips the orientation of $X$, the gray automorphism of both $X$ and $\overline X$.}
\label{fig:action_orientations}
\end{figure}

\heading{The Induced Action.}
Let $X$ be a permutation graph. The main difference is that both $X$ and $\overline{X}$ are
comparability graphs. By the results of Section~\ref{sec:comparability_graphs}, we know that
$\Aut(X)$ induces an action on both $\frako(X)$ and $\frako(\overline{X})$. We work with these two
actions as with one action on the pair $(\frako(X),\frako(\overline X))$, in other words on pairs
$(O,\overline O)$ such that $O \in \frako(X)$ and $\overline O \in \frako(\overline X)$.
Figure~\ref{fig:action_orientations} shows an example.

An action is called \emph{semiregular} if only the identity has a fixed point.
In other words, all stabilizers of a semiregular action are trivial.

\begin{lemma}\label{lem:action_semiregular}
The action of $\Aut(X)$ on $(\frako(X),\frako(\overline X))$ is semiregular.
\end{lemma}

\begin{proof}
We know that a permutation belonging to a stabilizer can only permute incomparable elements. Since
incomparable elements in $O$ are exactly the comparable elements in $\overline O$, the stabilizer is
trivial.\qed
\end{proof}

\begin{lemma}\label{lem:groups_prime_perm}
For a prime permutation graph $X$, $\Aut(X)$ is a subgroup of $\cyc_2^2$.
\end{lemma}

\begin{proof}
There are at most four pairs of orientations in $(\frako(X),\frako(\overline X))$, so by
Lemma~\ref{lem:action_semiregular} the order of $\Aut(X)$ is at most four.  If $\pi \in \Aut(X)$,
then $\pi^2$ fixes the orientation of both $X$ and $\overline X$. Therefore $\pi^2$ is an identity,
$\pi$ an involution and $\Aut(X)$ is a subgroup of $\cyc_2^2$.\qed
\end{proof}

Now, we are ready to characterize $\Aut(\perm)$:

\begin{proof}[Theorem~\ref{thm:perm_aut_groups}, sketch]
To show that $\Aut(\perm)$ is closed under (b) to (d), we use construction as in
Fig.~\ref{fig:counterexample}; see Appendix~\ref{sec:app_permutation_graphs}.

\begin{figure}[b]
\centering
\includegraphics[scale=0.8]{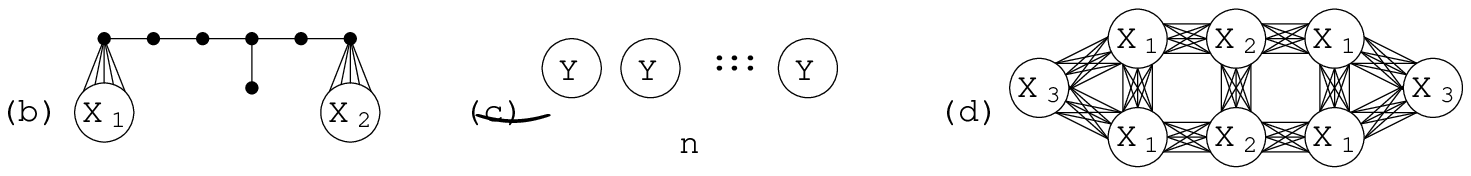}
\caption{The construction of the operations (b) to (d). It is easy to check that they are
permutation graphs with correct automorphism groups.}
\label{fig:counterexample}
\end{figure}

We apply the recursive procedure of Lemma~\ref{lem:recursive_aut_modular_tree}. We build $\Aut(T)$
recursively from the leaves to the root of $T$.  If the root node $R$ is degenerate, then we can
arbitrarily permute the isomorphic subtrees. Therefore, $\Aut(T)$ can be constructed using (b) and
(c). If the root node $R$ is a prime graph, we know that $\Aut(R)$ is by
Lemma~\ref{lem:groups_prime_perm} a subgroup of $\cyc_2^2$. Then $\Aut(T)$ can be constructed using
(d). See Appendix~\ref{sec:app_permutation_graphs} for details.\qed
\end{proof}

\heading{Geometry of Permutation Representations.}
We explain the result $\perm = \dim2$ of Even~\cite{even1972permutation}.  Let $O \in
\frako(X)$ and $\overline O \in \frako(\overline X)$, and let $\overline O_R$ be the reversal of
$\overline O$. We construct two linear orderings $L_1 = O \cup \overline O$ and $L_2 = O \cup
\overline O_R$. It follows that the comparable pairs in $L_1 \cap L_2$ are precisely the edges
$E(X)$.

\begin{figure}[t]
\centering
\includegraphics[scale=0.8]{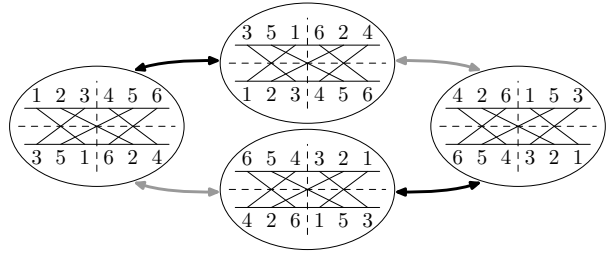}
\caption{Four representations of a symmetric permutation graph. The black automorphism is the
horizontal reflection with reverses $O$ and the gray automorphism is the vertical reflection which
reverses both $O$ and $\overline O$.}
\label{fig:action_representations}
\end{figure}

Consider a permutation representation of a symmetric prime permutation graph. The horizontal
reflection corresponds to exchanging $L_1$ and $L_2$, which is equivalent to reversing $\overline
O$. The vertical reflection corresponds to reversing both $L_1$ and $L_2$, which is equivalent to
reversing both $O$ and $\overline O$. The central rotation by $180^\circ$ is the combination of
both, which is equivalent to reversing $O$. See Fig.~\ref{fig:action_representations}. 

\section{$\boldsymbol k$-Dimensional Comparability Graphs} \label{sec:kdim}

We prove that $\Aut(\dim{4})$ contains all finite groups, i.e., each finite group can be realised as
an automorphism group of some $4$-dimensional comparability graph. Our construction also shows that
graph isomorphism testing of $\dim{4}$ is \cGI-complete.  Both results easily translate to $\dim{k}$
for $k > 4$ since $\dim{4} \subsetneq \dim{k}$.

\heading{The Construction.} Let $X$ be a graph with $V(X) = x_1, \dots, x_n$ and $E(X) = \{e_1,
\dots, e_m\}$. We define
$$P = \bigl\{p_i : x_i \in V(X)\bigr\},\qquad
  Q = \{q_{ik} : x_i \in e_k\},\qquad
  R = \bigl\{r_k : e_k \in E(X)\bigr\},$$
where $P$ represents the vertices, $R$ represents the edges and $Q$ represents the incidences
between the vertices and the edges.

\begin{figure}[b]
\centering
\includegraphics[scale=0.75]{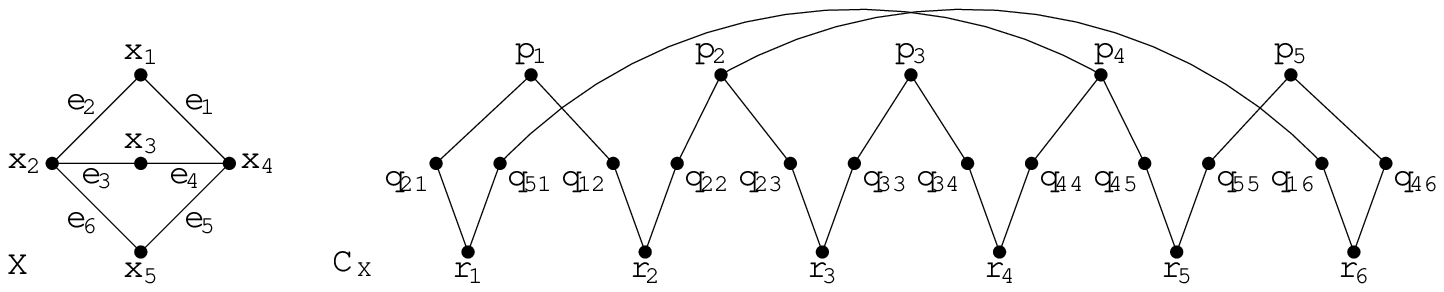}
\caption{The construction $C_X$ for the graph $X = K_{2,3}$.}
\label{fig:example_of_construction}
\end{figure}

The constructed comparability graph $C_X$ is defined as follows, see
Fig.~\ref{fig:example_of_construction}:
$$V(C_X) = P \cup Q \cup R,\qquad E(C_X) = \{p_iq_{ik}, q_{ik}r_k : x_i \in e_k\}.$$

\heading{Proof of Dimension 4.}
The harder part is to prove that the constructed graph $C_X$ has dimension four, which we can do
when $X$ is bipartite.

\begin{lemma} \label{lem:bipartite_4dim}
If $X$ is a connected bipartite graph, then $\dm(C_X) \leq 4$.
\end{lemma}

\begin{proof}
We construct four chains such that $L_1 \cap L_2 \cap L_3 \cap L_4$ have two vertices comparable if
and only if they are adjacent in $C_X$.  We describe linear chains as words containing each vertex
of $V(C_X)$ exactly once.  If $S_1,\dots,S_s$ is a sequence of strings, the symbol $\langle S_t :
\mathop\uparrow t \rangle$ is the concatenation $S_1S_2\dots S_s$ and $\langle S_t :
\mathop\downarrow t \rangle$ is the concatenation $S_sS_{s-1}\dots S_1$. When the arrows are omitted
as in $\langle S_t \rangle$, we concatenate in an arbitrary order.

First, we define the \emph{incidence string} $I_i$ which codes $p_i$ and its neighbors $q_{ik}$:
$$I_i = p_i \bigl\langle q_{ik} : p_iq_{ik} \in E(C_X) \bigr\rangle.$$
Notice that the concatenation $I_iI_j$ contains the right edges but it further contains edges
going from $p_i$ and $q_{ik}$ to $p_j$ and $q_{j\ell}$. We remove these edges by concatenation
$I_jI_i$ in some other chain. 

Since $X$ is bipartite, let $(A,B)$ be partition of its vertices. We define
\begin{eqnarray*}
P_A &=& \{p_i : x_i \in A\},\qquad \hskip 0.25em Q_A = \{q_{ik} : x_i \in A\},\\
P_B &=& \{p_j : x_j \in B\},\qquad Q_B = \{q_{jk} : x_j \in B\}.
\end{eqnarray*}
Each vertex $r_k$ has exactly one neighbor in $Q_A$ and exactly one in $Q_B$.

We construct the four chains as follows:
\begin{eqnarray*}
L_1 &=& \langle p_i : p_i \in P_A \rangle \langle r_kq_{ik} : q_{ik} \in Q_A, \mathop\uparrow k\rangle
\langle I_i : p_i \in P_B, \mathop\uparrow i\rangle,\\
L_2 &=& \langle p_i : p_i \in P_A \rangle \langle r_kq_{ik} : q_{ik} \in Q_A, \mathop\downarrow k\rangle
\langle I_i : p_i \in P_B, \mathop\downarrow i\rangle,\\
L_3 &=& \langle p_j : p_j \in P_B \rangle \langle r_kq_{jk} : q_{jk} \in Q_B, \mathop\uparrow k\rangle
\langle I_i : p_i \in P_A, \mathop\uparrow i\rangle,\\
L_4 &=& \langle p_j : p_j \in P_B \rangle \langle r_kq_{jk} : q_{jk} \in Q_B, \mathop\downarrow k\rangle
\langle I_i : p_i \in P_A, \mathop\downarrow i\rangle.
\end{eqnarray*}

\begin{figure}[t]
\centering
\includegraphics[scale=0.75]{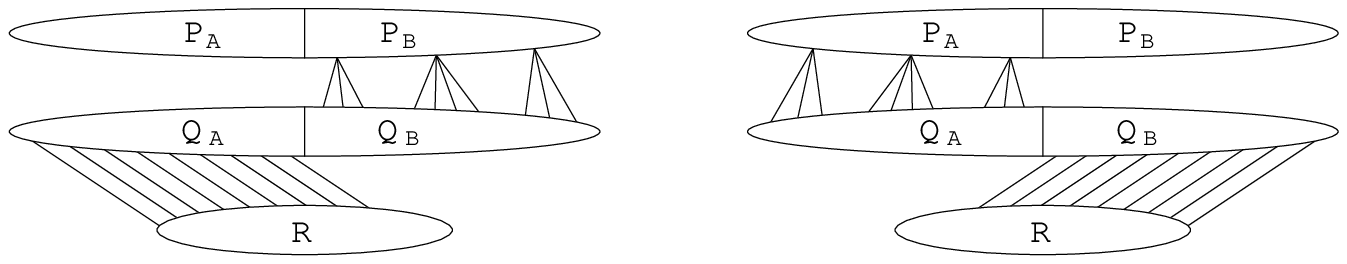}
\caption{On the left, the forced edges in $L_1 \cap L_2$, on the right in $L_3 \cap L_4$.}
\label{fig:proof_idea}
\end{figure}

See Fig.~\ref{fig:proof_idea} for properties of $L_1,\dots,L_4$. It is routine to verify
that the intersection $L_1 \cap L_2 \cap L_3 \cap L_4$ is correct; see Appendix~\ref{sec:app_proof_of_dim4}.\qed
\end{proof}

\begin{proof}[Theorem~\ref{thm:kdim_aut_groups_and_gi}]
It is sufficient to prove the statement for \dim4. Let $X$ be a connected graph with some
automorphism group $\Aut(X)$, and we assume that $X \not\cong C_n$. First, we take the bipartite
incidence graph $Y$ between $V(X)$ and $E(X)$, and it easily follows that $\Aut(Y) \cong \Aut(X)$.
Then we construct $C_Y$. In Appendix~\ref{sec:app_proof_of_dim4}, we have $\Aut(C_Y) \cong \Aut(Y)
\cong \Aut(X)$ and by Lemma~\ref{lem:bipartite_4dim}, we have that $C_Y \in \dim 4$. Similarly, if
two graphs $X_1$ and $X_2$ are given, we construct $C_{Y_1}$ and $C_{Y_2}$ such that $X_1 \cong X_2$
if and only if $C_{Y_1} \cong C_{Y_2}$; this gives the reduction which shows \cGI-completeness of
	graph isomorphism testing.\qed
\end{proof}

\section{Open Problems} \label{sec:conclusions}

We conclude with the following open problems:

\begin{problem}
What is $\Aut(\dim3)$?
\end{problem}

In Lemma~\ref{lem:groups_prime_perm}, we show that the automorphism group of a prime permutation
graph is always a subgroup of $\cyc_2^2$. Our proof does not give much structural insight into prime
permutation graphs.

\begin{problem}
What is the structure of prime permutation graphs? Can they be characterized?
\end{problem}

\bibliographystyle{plain}
\bibliography{automorphism_groups_of_comp}

\begin{thebibliography}{10}

\bibitem{babai1972automorphism}
L.~Babai.
\newblock Automorphism groups of planar graphs {II}.
\newblock In {\em Infinite and finite sets (Proc. Conf. Kestzthely, Hungary)},
  1973.

\bibitem{permutation_graphs}
K.~A. Baker, P.~C. Fishburn, and F.~S. Roberts.
\newblock Partial orders of dimension 2.
\newblock {\em Networks}, 2:11--28, 1972.

\bibitem{permutation_isomorphism}
C.~J. Colbourn.
\newblock On testing isomorphism of permutation graphs.
\newblock {\em Networks}, 11(1):13--21, 1981.

\bibitem{automorphism_algorithms_trees_int_planar}
C.~J. Colbourn and K.~S. Booth.
\newblock Linear times automorphism algorithms for trees, interval graphs, and
  planar graphs.
\newblock {\em SIAM J. Comput.}, 10(1):203--225, 1981.

\bibitem{dushnik_miller}
B.~Dushnik and E.~W. Miller.
\newblock Partially ordered sets.
\newblock {\em American Journal of Mathematics}, 63(3):600--610, 1941.

\bibitem{er63}
P.~Erd\H{o}s and A.~R\'enyi.
\newblock Asymmetric graphs.
\newblock {\em Acta Mathematica Academiae Scientiarum Hungarica},
  14(3--4):295--315, 1963.

\bibitem{even1972permutation}
S.~Even, A.~Pnueli, and A.~Lempel.
\newblock Permutation graphs and transitive graphs.
\newblock {\em Journal of the ACM (JACM)}, 19(3):400--410, 1972.

\bibitem{FKKN14}
J.~Fiala, P.~Klav\'{\i}k, J.~Kratochv\'{\i}l, and R.~Nedela.
\newblock Algorithmic aspects of regular graphs covers with applications to
  planar graphs.
\newblock In {\em ICALP 2014}, volume 8572 of {\em LNCS}, pages 489--501, 2014.

\bibitem{frucht1939herstellung}
R.~Frucht.
\newblock Herstellung von graphen mit vorgegebener abstrakter gruppe.
\newblock {\em Compositio Mathematica}, 6:239--250, 1939.

\bibitem{gallai1967transitiv}
T.~Gallai.
\newblock Transitiv orientierbare graphen.
\newblock {\em Acta Mathematica Hungarica}, 18(1):25--66, 1967.

\bibitem{golumbic1983comparability}
M.~C. Golumbic, D.~Rotem, and J.~Urrutia.
\newblock Comparability graphs and intersection graphs.
\newblock {\em Discrete Mathematics}, 43(1):37--46, 1983.

\bibitem{grohe_marx}
M.~Grohe and D.~Marx.
\newblock Structure theorem and isomorphism test for graphs with excluded
  topological subgraphs.
\newblock In {\em Proceedings of the Forty-fourth Annual ACM Symposium on
  Theory of Computing}, STOC '12, pages 173--192, 2012.

\bibitem{hsu1995m}
W.~L. Hsu.
\newblock {$\O(M \cdot N)$} algorithms for the recognition and isomorphism
  problems on circular-arc graphs.
\newblock {\em SIAM Journal on Computing}, 24(3):411--439, 1995.

\bibitem{jordan1869assemblages}
C.~Jordan.
\newblock Sur les assemblages de lignes.
\newblock {\em Journal f{\"u}r die reine und angewandte Mathematik},
  70:185--190, 1869.

\bibitem{kz}
P.~Klav\'{\i}k and P.~Zeman.
\newblock Automorphism groups of geometrically represented graphs.
\newblock In {\em STACS 2015}, volume~30 of {\em LIPIcs}, pages 540--553, 2015.

\bibitem{luks1982isomorphism}
E.~M. Luks.
\newblock Isomorphism of graphs of bounded valence can be tested in polynomial
  time.
\newblock {\em Journal of Computer and System Sciences}, 25(1):42--65, 1982.

\bibitem{mathon1979note}
R.~Mathon.
\newblock A note on the graph isomorphism counting problem.
\newblock {\em Information Processing Letters}, 8(3):131--136, 1979.

\bibitem{schoning1988graph}
U.~Sch{\"o}ning.
\newblock Graph isomorphism is in the low hierarchy.
\newblock {\em Journal of Computer and System Sciences}, 37(3):312--323, 1988.

\bibitem{yannakakis1982complexity}
Mihalis Yannakakis.
\newblock The complexity of the partial order dimension problem.
\newblock {\em SIAM Journal on Algebraic Discrete Methods}, 3(3):351--358,
  1982.

\end{thebibliography}

\newpage
\appendix

\section{Modular Trees} \label{sec:app_modular_trees}

The following lemma explains that $T$ encodes adjacencies in $X$:

\begin{lemma}\label{lem:adjacency_modular_tree}
We have $xy \in E(X)$ if and only if there exists an alternating path $xm_1m_2\dots m_ky$
in the modular tree $T$ such that each $m_i$ is a marker vertex and precisely the edges
$m_{2i-1}m_{2i}$ are tree edges.
\end{lemma}

\begin{proof}
Suppose that $xy \in E(X)$. If $xy \in E(T)$, then we are done. We assume that $xy \notin E(T)$.
The modular decomposition was constructed by a sequence of quotient operations. At some step of the
construction we get the last graph $X_0$ such that $xy \in E(X_0)$. Let $\calP$ be the modular
partition of $X_0$ chosen by the modular decomposition.  As in the construction of the modular tree,
we denote the marker vertices obtained from the contraction of the modules by $m_1, \dots, m_k$, and
the marker vertices attached to those by tree edges by $m_1', \dots, m_k'$.

We consider the next step of the modular decomposition. Suppose that $x \in M_i$ and $y \in M_j$. We
have that $x \in V(X_0[M_i])$ and $y \in V(X_0[M_j])$. From the construction of $T$, it follows
that $xm_i'$ and $ym_j'$ are normal edges and since $xy \in E(X_0)$, we also have that $m_im_j \in
E(X_0/\calP)$. The vertices $x m_i' m_i m_j m_j' y$ form an alternating path.

Now, we recursively construct an alternating path in $T$. From the construction of $T$, we have
that the vertices $x$ and $m_i'$ are connected by a normal edge.  Since the vertices $x$ and $m_i'$
are adjacent in the graph $X_0[M_i] \cup m_i'$, there exists an alternating path $P_i$ connecting
$x$ and $m_i'$ in the subtree of $T$ representing $X_0[M_i] \cup m_i'$. Similarly, we have an
alternating path $P_j$ connecting $y$ and $m_j'$ in some subtree of $T$ representing $X_0[M_j]
\cup m_j'$. The vertices $x P_i m_i' m_i m_j m_j' P_j y$ form a correct
alternating path in $T$.

The converse implication can be easily derived by reversing the process described above.\qed
\end{proof}

\begin{proof}[Lemma~\ref{lem:aut_group_modular_tree}]
First, we show that each automorphism $\sigma \in \Aut(T)$ induces a unique automorphism of $X$. We
define $\alpha = \sigma\restriction_A$. By Lemma~\ref{lem:adjacency_modular_tree} two vertices $x, y
\in V(X)$ are adjacent if and only if there exists and alternating path in $T$ connecting them.
Since $\sigma$ is an automorphism, we also have an alternating path between $\sigma(x)$ and
$\sigma(y)$. Therefore, $xy \in E(X) \iff \alpha(x)\alpha(y) \in E(X)$.

To obtain the converse implication, we prove that $\alpha \in \Aut(X)$ induces a unique automorphism
$\sigma \in \Aut(T)$. We define $\sigma(x) = \alpha(x)$ for a non-marker vertex $x$. On the marker
vertices, we define $\sigma$ recursively as follows. Let $\calP = \{M_1, \dots, M_k\}$ be a modular
partition of $X$ from the construction of the modular decomposition. It is easy to see that the
group $\Aut(X)$ induces a action on the partition $P$. If $\alpha(M_i) = M_j$, then clearly $X[M_i]$
and $X[M_j]$ are isomorphic. We define $\sigma(m_i) = m_j$ and $\sigma(m_i') = m_j'$, and finish the
rest recursively. Since $\sigma$ is an automorphism at each step of the construction, it follows
that $\sigma \in \Aut(T)$.\qed
\end{proof}

\begin{proof}[Lemma~\ref{lem:recursive_aut_modular_tree}]
We isomorphically label the vertices of isomorphic subtrees $T_i$. Each automorphism $\pi \in
\Aut(T)$ is a composition of two automorphisms $\sigma \cdot \tau$ where $\sigma$ maps each subtree
$T_i$ to itself, and $\tau$ permutes the subtrees as in $\pi$ while preserving the labeling.
Therefore, the automorphisms $\sigma$ can be bijectively identified with the elements of the direct
product $\Aut(T_1) \times \cdots \times \Aut(T_k)$ and the automorphisms $\tau$ with some element of
$\Aut(R)$.

Let $\pi,\pi' \in \Aut(T)$.  Consider the composition $\sigma \cdot \tau \cdot \sigma' \cdot \tau'$,
we want to swap $\tau$ with $\sigma'$ and rewrite this as a composition $\sigma \cdot \hat\sigma
\cdot \hat\tau \cdot \tau'$. Clearly the subtrees are permuted in $\pi \cdot \pi'$ exactly as in
$\tau \cdot \tau'$, so $\hat\tau = \tau$. On the other hand, $\hat\sigma$ is not necessarily equal
$\sigma'$. Let $\sigma'$ be identified with the vector
$$(\sigma_1',\dots,\sigma_k') \in \Aut(T_1) \times \cdots \times \Aut(T_k).$$
Since $\sigma'$ is applied after $\tau$, it acts on the subtrees permuted according to $\tau$.
Thus, $\hat\sigma$ is constructed from $\sigma$ by permuting the coordinates of its vector by
$\tau$:
$$\hat\sigma = (\sigma'_{\tau(1)},\dots,\sigma'_{\tau(k)}).$$
This is precisely the definition of the semidirect product.\qed
\end{proof}

\section{Permutation Graphs} \label{sec:app_permutation_graphs}

\begin{proof}[Theorem~\ref{thm:perm_aut_groups}]
Since $\{1\} \in \Aut(\perm)$, we need prove that $\Aut(\perm)$ is closed
under (b)--(d).
\begin{itemize}
\item
Let $G_1, G_2 \in \Aut(\perm)$, and let $X_1$ and $X_2$ be two permutation
graphs such that $\Aut(X_1) \cong G_1$ and $\Aut(X_2) \cong G_2$. We construct
a permutation graph $X$ by attaching $X_1$ and $X_2$ to an asymmetric
permutation graph; see Figure~\ref{fig:counterexample}b. Clearly, we get
$\Aut(X) \cong G_1 \times G_2$.
\item
Let $G \in \Aut(\perm)$, and let $Y$ be connected a permutation graph such that
$\Aut(Y) \cong G$. We construct a graph $X$ by taking the disjoint union of $n$
copies of $Y$; see Figure~\ref{fig:counterexample}. Clearly, we get $\Aut(X)
\cong G\wr \sym_n$.
\item
Let $G_1, G_2, G_3 \in \Aut(\perm)$, and let $X_1$, $X_2$, and $X_3$ be
permutation graphs such that $\Aut(X_i) \cong G_i$, for $i = 1, 2, 3$. We
construct a graph $X$ as shown in Figure~\ref{fig:counterexample}. Clearly, we
get $\Aut(X) \cong \bigl (G_1^4 \times G_2^2 \times G_3^2 \bigr) \rtimes \cyc_2^2$.
\end{itemize}

To show that for a given permutation graph $X$ the group $\Aut(X) \in
\Aut(\perm)$ we use Lemma~\ref{lem:recursive_aut_modular_tree}. Let $T$ be the
modular tree representing $X$, let $R$ be its root, and let $T_1, \dots, T_k$ be
the subtrees of $R$. By induction, we assume that $\Aut(T_i) \in \Aut(\perm)$,
and we show that also $\Aut(T) \in \Aut(\perm)$.  We distinguish two cases.
\begin{itemize}
\item
If $R$ is a degenerate node (an independent set or a complete graph), then
$\Aut(R)$ is a direct product of symmetric groups. By
Lemma~\ref{lem:recursive_aut_modular_tree}, we get
$$\Aut(T) \cong \bigl(\Aut(T_1) \times \cdots \times \Aut(T_k)\bigr) \rtimes \bigl (\sym_{\ell_1}
\times \cdots \times \sym_{\ell_m}\bigr),$$
where $\ell_1, \dots, \ell_m$ are the sizes of the isomorphism classes of $T_1, \dots, T_k$.
Let $G_{i}$ be the direct product of all $\Aut(T_j)$ such that $T_j$ belong to
the same isomorphism class $i$. We have
$$\Aut(T) \cong G_{1} \wr \sym_{\ell_1} \times \cdots \times G_{m}\wr
\sym_{\ell_m}.$$
Therefore $\Aut(X) \cong \Aut(T)$ can be constructed using (b) and (c) and it
belongs to $\Aut(\perm)$.
\item
If $R$ is a prime node, then by Lemma~\ref{lem:groups_prime_perm}, $\Aut(R)$ is
a subgroup of $\cyc_2^2$. The only interesting case is when $\Aut(R) \cong
\cyc_2^2$. From the orbit-stabilizer theorem, the action of $\cyc_2^2$ on $V(R)$
can have orbits of sizes $4$, $2$, and $1$. Moreover, each orbit of size $2$
corresponds to some stabilizer of size $2$. Since there are three subgroups of
$\cyc_2^2$ of size $2$, there can be possibly three types of orbits of size $2$.
By a geometric argument, we show that if $R$ is a prime permutation graph, then
one of the three subgroups of size $2$ can not be a stabilizer of any orbit of
size $2$, and therefore there are at most two types
of orbits of size $2$.

The non-identity elements $(1, 0)$, $(0, 1)$, and $(1, 1)$ of $\cyc_2^2$
correspond to the reflection $f$ of the permutation representation along the
vertical axis, reflection $f'$ along the horizontal axis, and rotation $r$
around the center by $180^\circ$, respectively; see
Figure~\ref{fig:action_representations}. The reflection $f$ stabilizes only
segments that that coincide with the vertical axis. Note that there can be at
most one such segment, since otherwise $R$ would not be prime.
Therefore, the reflection $f$ does not stabilize any orbit of size $2$.

Let $G_1$ be the direct product of all $\Aut(T_j)$ such that $T_j$ is attached
to a vertex of $R$ that belongs to an orbit of size four. The groups $G_2$ and
$G_3$ are defined similarly for the two types of orbits of size two, and $G_4$
for the orbits of size one. We have
$$\Aut(T) \cong \bigl (G_1^4 \times G_2^2 \times G_3^2 \times G_1\bigr)
\rtimes_{\varphi}
\cyc_2^2 \cong \bigl (G_1^4 \times G_2^2 \times G_3^2\bigr ) \rtimes \cyc_2^2 \times
G_4,$$
where $\varphi \colon \cyc_2^2 \to \Aut \bigl ( G_1^4 \times G_2^2 \times G_3^2
\times G_1 \bigr )$ is the homomorphism defined as follows. The automorphism
$\varphi(1,0)$ swaps the first two components of $G_1^4$, swaps the components of
of $G_2^2$, fixes the components of $G_3^2$, and fixes $G_1$. The automorphism
$\varphi(0, 1)$ swaps the second two components of $G_1^4$, fixes the components
of $G_2^2$, swaps the components of $G_3^2$, and fixes $G_1$.
We get that $\Aut(X) \cong \Aut(T)$ can be constructed using (b) and (d) and it
belongs to $\Aut(\perm)$.
\end{itemize}
\end{proof}

\section{$\boldsymbol{k}$-dimensional Comparability Graphs} \label{sec:app_proof_of_dim4}

\begin{lemma} \label{lem:construction_preserves_aut_group}
Let $X$ be a connected graph such that $X \not\cong C_n$. Then $$\Aut(C_X) \cong \Aut(X).$$
\end{lemma}

\begin{proof}
All vertices of $Q$ and $R$ have degree two, and by our assumption at least one vertex $p_i$ in $P$ has a
different degree. Therefore, we obtain $P$ as the set of the vertices in $C_X$ whose distance from
$p_i$ is divisible by four, $Q$ as the set of their neighbors and $R$ as the remaining vertices.
Every automorphism of $C_X$ has to preserve this partition, therefore it induces an automorphism of
$X$. Since this construction does not depend on the labeling, every automorphism of $X$ is an
automorphism of $C_X$. Therefore, $\Aut(C_X) \cong \Aut(X)$.\qed
\end{proof}

\begin{proof}
We conclude the proof of Lemma~\ref{lem:bipartite_4dim} by verify the construction:
\begin{eqnarray*}
L_1 &=& \langle p_i : p_i \in P_A \rangle \langle r_kq_{ik} : q_{ik} \in Q_A, \mathop\uparrow k\rangle
\langle I_i : p_i \in P_B, \mathop\uparrow i\rangle,\\
L_2 &=& \langle p_i : p_i \in P_A \rangle \langle r_kq_{ik} : q_{ik} \in Q_A, \mathop\downarrow k\rangle
\langle I_i : p_i \in P_B, \mathop\downarrow i\rangle,\\
L_3 &=& \langle p_j : p_j \in P_B \rangle \langle r_kq_{jk} : q_{jk} \in Q_B, \mathop\uparrow k\rangle
\langle I_i : p_i \in P_A, \mathop\uparrow i\rangle,\\
L_4 &=& \langle p_j : p_j \in P_B \rangle \langle r_kq_{jk} : q_{jk} \in Q_B, \mathop\downarrow k\rangle
\langle I_i : p_i \in P_A, \mathop\downarrow i\rangle.
\end{eqnarray*}

The four defined chains have the following properties, see Figure~\ref{fig:proof_idea}:

\begin{packed_itemize}
\item The intersection $L_1 \cap L_2$ forces the correct edges between $Q_A$ and $R$ and between
$P_B$ and $Q_B$. It poses no restrictions between $Q_B$ and $R$ and between $P_A$ and the rest of
the graph.
\item Similarly the intersection $L_3 \cap L_4$ forces the correct edges between $Q_B$ and $R$ and
between $P_A$ and $Q_A$. It poses no restrictions between $Q_A$ and $R$ and between $P_B$ and the
rest of the graph.
\end{packed_itemize}

\medskip
\noindent\emph{Claim 1: The edges in $Q \cup R$ are correct.} For every $k$, we get $r_k$ adjacent to both
$q_{ik}$ and $q_{jk}$ since it appear on the left in $L_1,\dots,L_4$. On the other hand,
$q_{ik}q_{jk} \notin E(C_X)$ since they are ordered differently in $L_1$ and $L_3$.

For every $k < \ell$, there are no edges between
$N[r_k] = \{r_k,q_{ik},q_{jk}\}$ and $N[r_\ell] = \{r_\ell,q_{s\ell},q_{t\ell}\}$. This can be shown
by checking the four ordering of these six elements:
\begin{eqnarray*}
\text{in $L_1$:}\quad r_k q_{ik}\, \boxed{r_\ell q_{s\ell}}\, q_{jk}\, \boxed{q_{t \ell}},\quad&&\quad
\text{in $L_2$:}\quad \boxed{r_\ell q_{s\ell}}\, r_k q_{ik} q_{jk}\, \boxed{q_{t \ell}},\\
\text{in $L_3$:}\quad r_k q_{jk}\, \boxed{r_\ell q_{t\ell}}\, q_{ik}\, \boxed{q_{s \ell}},\quad&&\quad
\text{in $L_4$:}\quad \boxed{r_\ell q_{t\ell}}\, r_k q_{jk} q_{ik}\, \boxed{q_{s \ell}},
\end{eqnarray*}
where the elements of $N[r_\ell]$ are boxed.\claimqed

\medskip
\noindent\emph{Claim 2: The edges in $P$ are correct.}
We show that there are no edges between $p_i$ and $p_j$ for $i \ne j$ as follows.  If both belong to
$P_A$ (respectively $P_B$), then they are ordered differently in $L_3$ and $L_4$ (respectively $L_1$
and $L_2$). If one belongs to $P_A$ and the other one to $P_B$, then they are ordered differently in
$L_1$ and $L_3$.\claimqed

\medskip
\noindent\emph{Claim 3: The edges between $P$ and $Q \cup R$ are correct.}
For every $p_i \in P$ and $r_k \in R$, we have $p_ir_k \notin E(C_X)$ because they are
ordered differently in $L_1$ and $L_3$. On the other hand, $p_iq_{ik} \in E(C_X)$, because $p_i$ is
before $q_{ik}$ in $I_i$, and for $p_i \in P_A$ in $L_1$ and $L_2$, and for $p_i \in
B$ in $L_3$ and $L_4$.

It remains to show that $p_iq_{jk} \notin E(C_X)$ for $i \ne j$.  If both $p_i$ and $p_j$ belong to
$P_A$ (respectively $P_B$), then $p_i$ and $q_{jk}$ are ordered differently in $L_3$ and $L_4$
(respectively $L_1$ and $L_2$). And if one belongs to $P_A$ and the other one to $P_B$, then $p_i$
and $q_{jk}$ are ordered differently in $L_1$ and $L_3$.\claimqed

These three claims show that comparable pairs in the intersection $L_1 \cap L_2 \cap L_3 \cap L_4$
are exactly the edges of $C_X$, so $C_X$ is a comparability graph with the dimension at most
four.\qed
\end{proof}

\end{document}